\documentclass{elsart3p}

\usepackage{epsfig}

\journal{Physics Letters B}
\begin{document}
\begin{frontmatter}
\title{Study of dielectron production in C+C collisions at 1~AGeV}


\collaboration{HADES Collaboration}

\author[8]{G.~Agakishiev},
\author[1]{C.~Agodi},
\author[15]{H.~Alvarez-Pol},
\author[3,e]{A.~Balanda},
\author[9]{R.~Bassini},
\author[1,a]{G.~Bellia},
\author[15]{D.~Belver},
\author[6]{A.~Belyaev},
\author[2]{A.~Blanco},
\author[11]{M.~B\"{o}hmer},
% ...

\vspace*{0.3cm}

\address[1]{Istituto Nazionale di Fisica Nucleare - Laboratori Nazionali del Sud, 95125~Catania, Italy}
\address[2]{LIP-Laborat\'{o}rio de Instrumenta\c{c}\~{a}o e F\'{\i}sica Experimental de Part\'{\i}culas,
3004-516~Coimbra, Portugal}
\address[3]{Smoluchowski Institute of Physics, Jagiellonian University of Cracow, 30-059~Krak\'{o}w, Poland}
\address[4]{Gesellschaft f\"{u}r Schwerionenforschung mbH, 64291~Darmstadt, Germany}
\address[5]{Institut f\"{u}r Strahlenphysik, Forschungszentrum Dresden-Rossendorf, 01314~Dresden, Germany}
\address[6]{Joint Institute of Nuclear Research, 141980~Dubna, Russia}
\address[7]{Institut f\"{u}r Kernphysik, Johann Wolfgang Goethe-Universit\"{a}t, 60438 ~Frankfurt, Germany}
\address[8]{II.Physikalisches Institut, Justus Liebig Universit\"{a}t Giessen, 35392~Giessen, Germany}
\address[9]{Istituto Nazionale di Fisica Nucleare, Sezione di Milano, 20133~Milano, Italy}
\address[10]{Institute for Nuclear Research, Russian Academy of Science, 117312~Moscow, Russia}
\address[11]{Physik Department E12, Technische Universit\"{a}t M\"{u}nchen, 85748~M\"{u}nchen, Germany}
\address[12]{Department of Physics, University of Cyprus, 1678~Nicosia, Cyprus}
\address[13]{Institut de Physique Nucl\'{e}aire (UMR 8608), CNRS/IN2P3 - Université Paris Sud,
F-91406~Orsay Cedex, France}
\address[14]{Nuclear Physics Institute, Academy of Sciences of Czech Republic, 25068~Rez, Czech Republic}
\address[15]{Departamento de F\'{\i}sica de Part\'{\i}culas, University of Santiago de Compostela,
15782~Santiago de Compostela, Spain}
\address[16]{Instituto de F\'{\i}sica Corpuscular, Universidad de Valencia-CSIC, 46971~Valencia, Spain}

\vspace*{0.3cm}

\address[a]{Also at Dipartimento di Fisica e Astronomia, Universit\`{a} di Catania, 95125~Catania, Italy}
\address[b]{Also at ISEC Coimbra, ~Coimbra, Portugal}
\address[c]{Also at Technische Universit\"{a}t Dresden, 01062~Dresden, Germany}
\address[d]{Also at Dipartimento di Fisica, Universit\`{a} di Milano, 20133~Milano, Italy}
\address[e]{Also at Panstwowa Wyzsza Szkola Zawodowa , 33-300~Nowy Sacz, Poland}
\corauth[cor]{Corresponding author.}

\begin{abstract}
The emission of $e^+ e^-$ pairs from C+C collisions
at an incident energy of 1 GeV per nucleon has been investigated.
The measured production probabilities, spanning from the $\pi^0$-Dalitz
to the $\rho/\omega$ invariant-mass region, display a strong excess above
the cocktail of standard hadronic sources.  The bombarding-energy dependence
of this excess is found to scale like pion production, rather than like eta
production.  The data are in good agreement with results obtained in the
former DLS experiment.
\end{abstract}

\end{frontmatter}

\section{Introduction}

An enhanced yield of dileptons with masses below the vector-meson (i.e. $\rho^0$ and
$\omega$ meson) pole mass appears to be a general feature of heavy-ion reactions, from
bombarding energies as low as 1~AGeV, studied by the former DLS experiment \cite{dls1} at the
Bevalac, through the range of SPS energies (40--158~AGeV) used by the CERES \cite{na45}
and NA60 \cite{na60} experiments at CERN, up to the highest energies
(with $\sqrt{s}_{NN} = 200$~GeV) employed by the PHENIX experiment \cite{phenix}
at the RHIC collider.  This enhancement is defined as the excess of the
measured pair yield over the summed-up cocktail of dileptons from long-lived sources,
namely the decays of $\pi^0$, $\eta$, and $\omega$ mesons.  It is hence expected to
probe the early phase of the collision and, in particular, the in-medium behavior
of short-lived hadronic resonances, as e.g. the $\rho$ and $\Delta$ \cite{rapp,weise,rhospectral}.
However, while the dilepton enhancement observed at the SPS has been related to
modifications of the $\rho$-meson spectral function in the hadronic medium \cite{vanhees,renk},
the large pair yields found by DLS in 1~AGeV C+C and Ca+Ca collisions remain to
be explained satisfactorily \cite{ernst,CassingBrat,fuchs,cozma}.

The High-Acceptance DiElectron Spectrometer HADES at GSI, Darmstadt, has started a
systematic investigation of dilepton production in the SIS/Bevalac energy regime of
1--2 AGeV.  First results obtained in 2~AGeV C+C collisions \cite{hades1} confirmed indeed
the general observation of enhanced emission of e$^+$e$^-$ pairs with invariant masses
of 0.15 -- 0.50~GeV/c$^2$.  In this Letter we report on a measurement of inclusive
electron-pair emission from $^{12}$C+$^{nat}$C collisions at a kinetic beam energy of 1~AGeV,
i.e. the energy of the DLS experiment.  Together with our results obtained at 2~AGeV,
this allows to discuss the beam-energy dependence of the pair yield.  In addition, a
direct comparison with the DLS results \cite{dls1} becomes now possible.

\section{Experiment}

In the experiment, a $^{12}$C beam of $10^6$ particles/s was incident on
a target of natural carbon with a thickness corresponding to $4.6\%$ of one nuclear
interaction length.  The configuration of the HADES spectrometer, described
in detail in Refs. \cite{HADESNIM,hades2}, was basically identical to the one
used in our former 2~AGeV C+C run \cite{hades1}.  To increase the acquired
pair statistics, besides a charged-particle multiplicity trigger (LVL1),
an online electron identification (LVL2) has been operated as part of the two-level
trigger system \cite{hades1}.  All results presented here were obtained from
events with a positive LVL2 trigger decision, with a total statistics corresponding
to $1.1\times10^9$ LVL1 triggered events.  One major difference to the former run
was the presence of up to four planes of tracking drift chambers (two inner and one or
two outer), of which, however, only the two inner planes were used in the extraction of
the results presented here.  This modus operandi was motivated by the goal to
simplify any comparison with the results of our 2~AGeV run obtained within a
low-resolution mode.  In the latter the reconstruction of outer track segments is based
solely on the position information obtained from the Time-of-Flight and Pre-Shower
detectors (see \cite{hades1} for details).  The lepton identification and
dilepton-reconstruction were done following the same scheme as used for the
2~AGeV data.  All this resulted in a momentum and mass resolution, as well as
a pair acceptance very similar to those characteristic of the former run,
making the comparison of the two data sets unproblematic.  As will become apparent
below, the achieved mass resolution of $\sigma_{M_{ee}}/M_{ee}=9\%$ at
$M_{ee}=0.8$ GeV/c$^2$ is largely sufficient to resolve all spectral features.
Results from a high-resolution analysis based on all four drift-chamber planes
will, however, be presented in another, forthcoming publication.

In the pair analysis \cite{pachmayer}, opposite-sign $e^+e^-$, as well as like-sign
$e^+e^+$ and $e^-e^-$ pairs were formed and subjected to common selection
criteria, in particular to an opening-angle cut of $\theta_{ee}>9^{\circ}$.
From the reconstructed like-sign distributions, e.g. the invariant mass
$dN^{++}/dM_{ee}$ and $dN^{--}/dM_{ee}$, the combinatorial background (CB)
of uncorrelated pairs was calculated bin by bin as $N_{CB} = 2\sqrt{N^{++}N^{--}}$.
For masses $M_{ee}>0.2$ GeV/$c^2$, where statistics is small, the CB was
obtained by an event-mixing procedure.  Finally, after subtracting the CB,
a total of $\simeq 18000$ signal pairs ($\simeq650$ with $M_{ee}>0.15$ GeV/c$^2$)
was thus reconstructed.

Detector and reconstruction efficiencies $\epsilon_{\pm}(p,\theta,\phi)$ were
determined from Monte-Carlo simulations by embedding electron and positron tracks into
$^{12}$C+$^{12}$C events generated with the UrQMD transport model \cite{bleicher}.
The experimental data were then corrected on a pair-by-pair basis with the weighting
factor $1/E_{+-}$, with $E_{+-}=\epsilon_+ \cdot \epsilon_-$.  The geometrical pair
acceptance of the HADES detector was obtained in analogy to the pair efficiency
as the product of two single-electron acceptances $A_{\pm}(p,\theta,\phi)$.
The resulting matrices, together with a momentum resolution function,
constitute the HADES acceptance filter (for more details see \cite{hades1}).
These acceptance matrices are available from the authors on request.

\section{Results on pair production}

\begin{figure}[htb]

\mbox{\epsfig{figure={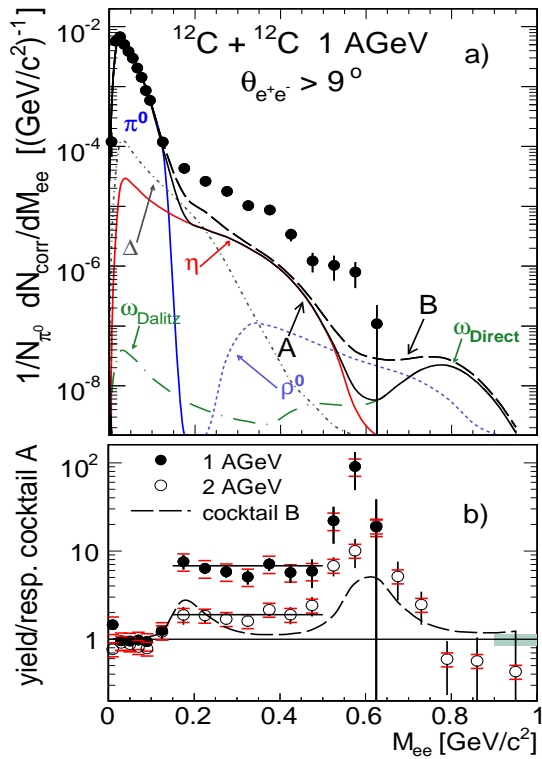}, width=1.0\linewidth,height=1.3\linewidth}}
\vspace*{-0.5cm}

  \caption[]{
  (a) Normalized, background-subtracted and efficiency-corrected $e^+e^-$
  invariant-mass distribution compared to thermal dielectron cocktails of free
  $\pi^0$, $\eta$ and $\omega$ decays (cocktail A, solid line), as well as
  including $\rho$ and $\Delta$ resonance decays (cocktail B, long-dashed line).
  Only statistical errors are shown.
  (b) Ratio of data and cocktail A (full symbols), compared to the corresponding
  ratio from the 2~AGeV C+C run \cite{hades1} (open symbols).
  Statistical and systematic errors of the measurement are shown as vertical
  and horizontal bars, respectively;  the shaded area depicts the 15\%
  normalization error.  Averages over the 0.15--0.50 GeV/c$^2$ mass range are
  indicated by lines; they correspond to the $F$ factors discussed in the text.
  The dashed curve corresponds to the ratio cocktail~B/cocktail~A for 1~AGeV.}

  \label{signal}

\end{figure}

Fig.~\ref{signal}(a) shows the $e^+e^-$ invariant-mass distribution of the
signal pairs after efficiency correction and normalized to the average
number of charged pions $N_{\pi}=\frac{1}{2}(N_{\pi^+}+N_{\pi^-})$,
measured as well with HADES and extra\-polated to 4$\pi$ solid angle.
In the isospin-symmetric system $^{12}$C+$^{12}$C, and for small contributions
from $\eta$ and $\omega$ decays, $N_{\pi}$ is also a good measure of the
$\pi^0$ yield, i.e.  $N_{\pi}$ = $N_{\pi^0}$.  This way of normalizing the pair
spectra compensates to first order the bias caused by the implicit centrality
selection of our trigger. Indeed, simulations based on UrQMD events show that
LVL1 events have an average number of participating nucleons $A_{part}=8.6$,
instead of 6 for true minimum-bias events.  The pion multiplicity per number
of participating nucleons $M_{\pi}/A_{part}=0.061\pm0.009$ obtained in our
experiment agrees with previous measurements of charged and neutral pions
\cite{taps1,kaos} within the quoted error of $15\%$.  The latter is
dominated by systematic uncertainties in the acceptance and efficiency
corrections of the charged-pion analysis, and it represents our
overall normalization error.  In addition, the uncertainties
caused by the lepton-efficiency correction and the CB subtraction
add up quadratically to point-to-point systematic errors of $22\%$ on
$dN^{+-}/dM_{ee}$.

In Fig.\ref{signal} we compare the data with a pair cocktail calculated
from free $\pi^0 \rightarrow \gamma e^{+}e^{-}$, $\eta \rightarrow \gamma e^{+}e^{-}$,
$\omega  \rightarrow \pi^{0}e^{+}e^{-}$, and $\omega \rightarrow e^{+}e^{-}$
meson decays only (cocktail A).  This cocktail aims at representing all
contributions from decays of mesons in vacuum after the chemical and thermal
freeze-out of the fireball.  While the first two of these sources are directly
constrained by published data \cite{taps1} with uncertainties of
$10\%$ ($\pi^0$) and $25\%$ ($\eta$), respectively, the (small) multiplicity of
the $\omega$ meson is taken from an $m_{\perp}$-scaling ansatz \cite{bratkovskaya1}.
We followed here the same procedure as applied for our 2~AGeV data \cite{hades1},
making use of the PLUTO event generator \cite{PLUTO}.  Meson production was modeled
assuming emission from a thermal source with an effective temperature $T=55$~MeV;
C+C being a very small system, zero radial expansion ($\beta_r$=0) has been assumed.
Furthermore, for $\pi^0$ and $\eta$ mesons, an anisotropic polar distribution of
the type $dN/d\cos(\theta_{CM}) \sim 1+a_2 \cdot cos^2(\theta_{CM})$ was used:
with $a_2=0.5$ for $\pi^0$, consistent with our charged-pion analysis, and $a_2=0.8$
for $\eta$, as inspired by transport calculations \cite{bratkovskaya1,bleicher}.
Effects due to a possible polarization of the virtual photon \cite{bratkovskaya2}
were not considered.  For the $\omega$ we have simply assumed an isotropic
decay pattern.  The accepted $\pi^0$ ($\eta$) Dalitz yield was found to change
by less than 15\% (10\%) when varying the source parameters over a broad range,
namely 0 -- 0.3 for $\beta_r$, 0 -- 0.8 for $a_2$, and 40 -- 70 MeV for $T$.
The cocktail does hence not depend much on our particular choice of these
parameters.

Whereas experimental data and simulated cocktail A (solid line in
Fig.~\ref{signal}(a) ) are in good agreement in the $\pi^0$ region, the
cocktail strongly undershoots the data for $M_{ee}>0.15$ GeV/c$^2$, much
more so than in our previous 2~AGeV data set, clearly calling for additional
dilepton sources.  This conclusion had also been reached by the authors of
Ref. \cite{taps2} who showed that their data on $\eta$ production
fix the contribution of $\eta$ Dalitz decays to pair production.
Additional contributions are of course expected from the decay of short-lived
resonances, mainly the $\Delta(1232)$ and the $\rho$, excited in the
early phase of the collision, and we have made an attempt to account for
them, schematically at least,  in our simulation as well.  To include in the
dilepton cocktail pairs from $\Delta^{0,+}\to Ne^+e^-$ decays, we assumed that
in the beam-energy regime of 1--2 AGeV the $\Delta$ yield scales with the
$\pi^0$ yield measured at freeze-out and we employed the $\Delta$ decay rate
calculated in \cite{ernst}.  To determine the $\rho$-meson contribution, we used a
similar prescription as for the $\omega$.  For this broad resonance
($\Gamma_{\circ} = 0.15$ GeV/c$^2$), described in our simulation by a relativistic
Breit-Wigner function with mass-dependent width (following \cite{CassingBrat}),
$m_{\perp}$ scaling, as well as the additional $1/M^3$ dependence of
$\Gamma(M)_{\rho \rightarrow e^+e^-}$
(imposed by vector dominance \cite{VDM}) strongly enhance the low-mass tail,
resulting in the skewed spectral shape visible in Fig.~\ref{signal}(a).  Finally,
Dalitz contributions from the heavier baryon resonances (N$^*$(1520), N$^*$(1535), etc.)
turned out to be negligeable in our thermal framework.  The full cocktail thus
generated (cocktail B) is shown in the figure as a long-dashed line.  One can see
that the simulated yield increases somewhat above 0.15 GeV/c$^2$, but obviously our
second calculation also remains far from reproducing the data.

The features of the dielectron mass spectrum are to some extent obscured by its
very steep fall, and in order to take this global trend out and make the
characteristics of the excess pair yield more visible, we display
in Fig.~\ref{signal}(b) the ratio of the data and cocktail A.  This
ratio is basically unity at low masses, dominated by the $\pi^0$ Dalitz pairs,
but above $M$=0.15 GeV/c$^2$ it is very much larger, indicating the onset of
processes not accounted for in our simple-minded cocktail calculation.
Fig.~\ref{signal}(b) also shows the corresponding ratio obtained from our previous
2~AGeV measurement \cite{hades1}, i.e. dividing those data by their respective
cocktail~A.  As already pointed out above, it is evident that at 1~AGeV the overshoot
of the data is much stronger than at 2~AGeV.  To quantify this behavior in the plateau
region of $M = 0.15 - 0.50$ GeV/c$^2$, we define for this mass range an average
enhancement above the known $\eta$ Dalitz contribution as $F = Y_{tot}/Y_{\eta}$.
This ratio, indicated in the figure by horizontal lines, amounts to
$F(1.0) = Y_{tot}(1.0)/Y_{\eta}(1.0) = 6.8 \pm 0.6 (stat) \pm 1.3 (sys) \pm 2.0 (\eta)$
at 1~AGeV, and to $F(2.0) = 1.9 \pm 0.2 (stat) \pm 0.3 (sys) \pm 0.3 (\eta)$
at 2~AGeV.  The third error, labeled ($\eta$), gives the uncertainty caused by
the quoted errors on the $\eta$ multiplicities.  Due to a re-evaluation of our pion
normalization, the value given here for $F(2.0)$ is by 10\% lower than the one cited
in \cite{hades1}.  Assuming now that the excess pairs have in this mass region an
overall acceptance close to that of $\eta$ Dalitz pairs, one can also compare $F(1.0)$
to the enhancement factor observed in C+C by DLS at a beam energy of
1.04~AGeV \cite{dls1}.  As the HADES and DLS geometric acceptances do not
fully overlap (see discussion below), this assumption is indeed necessary to make a
meaningful comparison of the ratios obtained within the respective acceptances.
Using the DLS data and a DLS-filtered PLUTO cocktail generated for 1.04~AGeV, we obtain
a factor of $F(1.04) = 6.5 \pm 0.5 (stat) \pm 2.1 (sys) \pm 1.5 (\eta)$.  The DLS result
is hence in good agreement with our 1~AGeV measurement.  Table~\ref{tab1} summarizes
all pair excess factors, together with their uncertainties.

\begin{table}[htb]
  \caption[]{Dielectron excess factor ($F$) and inclusive excess multiplicity ($N_{exc}$)
             from $^{12}$C+$^{nat}$C in the pair mass range $M = 0.15 - 0.50$ GeV/c$^2$
             as a function of bombarding energy ($E_b$).  Errors given from left to right are
             statistical (1), systematic (2), and due to the $\eta$ multiplicity (3).}
  \vspace*{0.2cm}
  \begin{center}
  \begin{tabular}{ c c c }
  \hline
  $~E_{b}$ [AGeV]~ & ~$F = Y_{tot}/Y_{\eta}$~ & ~$N_{exc}$ [10$^{-6}$]~\\
  \hline
  1.0  & $ ~6.8 \pm 0.6 \pm 1.3 \pm 2.0~ $ & $ ~6.8 \pm 0.7 \pm 1.5 \pm 0.3~ $\\
  2.0  & $ ~1.9 \pm 0.2 \pm 0.3 \pm 0.3~ $ & $ ~18 \pm 4 \pm 7 \pm 4~ $\\
  1.04 & $ ~6.5 \pm 0.5 \pm 2.1 \pm 1.5~ $ & $ ~8.4 \pm 0.7 \pm 3.4 \pm 0.3~ $\\
  \hline
  \end{tabular}
  \end{center}
  \label{tab1}
\end{table}

Continuing the discussion of Fig.~\ref{signal}(b), at masses above 0.50 GeV/c$^2$
the ratio of data and cocktail A develops for both beam energies a pronounced
maximum around $M_{ee}\sim 0.60$ GeV/c$^2$.  This is mainly due to the lack
of yield in the cocktail at these masses and can hence be considered to be an
artifact of the chosen way of presenting our data.  On the other hand, the mass
region between the $\eta$ and the $\omega$ pole is expected to be dominated by the
low-mass tail of the broad $\rho$ resonance, whose population is favored by the
available phase space.  Cocktail B, which includes $\rho$ and $\Delta$ decays,
shows indeed some enhancement with respect to cocktail A, as visible in
Fig.~\ref{signal}(b), but also at large masses it lies far below the observed
pair yield.

\section{Evolution with bombarding energy}

\begin{figure}[htb]

\mbox{\epsfig{figure={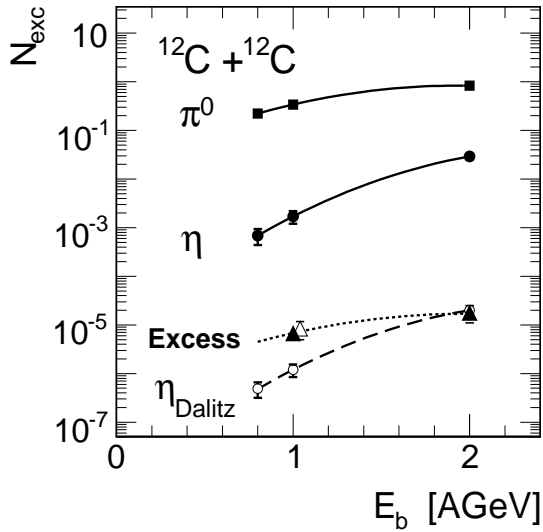}, width=1.0\linewidth,height=1.10\linewidth}}
\vspace*{-0.7cm}

  \caption[]{
  Inclusive multiplicity of the pair excess ($N_{exc}$) in the mass range
  $M_{ee} = 0.15 - 0.50$ GeV/c$^2$ as function of bombarding energy $E_b$
  (full triangles: HADES, open triangle: DLS).
  Also shown are the $\pi^0$ and $\eta$ inclusive multiplicities in $^{12}$C+$^{nat}$C
  collisions \cite{taps1} (full squares and circles), as well as the corresponding
  $\eta$ Dalitz decay contribution (open circles) summed over $M_{ee} = 0.15 - 0.50$
  GeV/c$^2$.  For comparison with $N_{exc}$, scaled-down $\pi^0$ (dotted) and absolute
  $\eta$ Dalitz (dashed) curves are shown (see discussion in the text).
  }

  \label{excitation}

\end{figure}

It is interesting to compare the beam-energy dependence of the pair
excess with that of neutral meson production in the C+C system.
The latter has been investigated systematically with photon calorimetry
\cite{taps1, taps2}, revealing that between 1 and 2~AGeV inclusive $\pi^0$
and $\eta$ multiplicities increase by factors $N_{\pi}(2.0)/N_{\pi}(1.0)=2.4\pm0.3$
and $N_{\eta}(2.0)/N_{\eta}(1.0)=17\pm 5$, respectively.  The excitation functions
of inclusive $\pi^0$ and $\eta$ production are shown in Fig.~\ref{excitation}, together
with the corresponding pair multiplicity from $\eta$ Dalitz decays (with
BR(${\eta \rightarrow \gamma e^+ e^-}$) = 0.6\%) within the 0.15-0.50 GeV/c$^2$
mass range, amounting to 11.6\% of all such pairs.  The energy scaling of the
inclusive excess pair multiplicity, i.e. for full solid angle and minimum bias,
($N_{exc}$) has been obtained from the measured excess factors $F(2.0)$ and $F(1.0)$
and the corresponding inclusive, i.e. mimimum bias $\eta$ multiplicities
(assuming similar detector acceptances for $\eta$ Dalitz and excess pairs) in the
following way: $N_{exc} = N_{tot}-N_{\eta} = (F-1) \; N_{\eta}$.
The resulting excess multiplicities are depicted in Fig.~\ref{excitation} for both
energies studied with HADES, together with the 1.04~AGeV DLS point; they are also
listed with their uncertainties in Table~\ref{tab1}.  From our data it follows that
$N_{exc}(2.0)/N_{exc}(1.0)=2.6\pm 0.6 (stat) \pm 0.6 (sys) \pm 0.5 (\eta)$.  This
energy dependence is hence remarkably similar to the evoluation of pion production, but
very different from that of $\eta$ production.  It is also apparent in Fig.~\ref{excitation}
in the direct comparison of the excess with the pion multiplicity scaled to hit the
2~AGeV point (dotted line), as well as with the absolute eta Dalitz contribution (dashed line).
This surprising behavior suggests that, at the bombarding energies discussed here,
the pair excess is not driven by the excitation of heavy resonances, but rather by
low-energy processes, like pion production and propagation, involving e.g. $\Delta$
and low-mass $\rho$ excitations, and possibly bremsstrahlung processes.

Recent one-boson exchange (OBE) calculations \cite{mosel,kaptari} of nucleon-nucleon
bremsstrahlung continue and extend the realistic treatment of interactions in the pp
channel of \cite{deJong}.  They indicate a much more important role of bremsstrahlung
in low-mass dilepton production than hitherto suspected.  According to these models,
in nucleon-nucleon collisions the quasi-elastic bremsstrahlung contribution is of
similar size as the $\Delta$ Dalitz part, accounting together for the total dilepton
yield above the $\pi^0$ Dalitz peak at beam energies below the $\eta$ production threshold
($E_{th} = 1.27$ GeV).  The OBE calculations also show that a consistent treatment of
both processes is not straightforward.  Conclusions on their validity can hence be
drawn only once more results on the elementary reactions pp and dp will become
available \cite{hades3}.

\section{Comparison with DLS}

We have already shown above that our pair excess at 1~AGeV agrees well with the
DLS measurement performed at 1.04~AGeV.  In the last section of this Letter we want
to present, however, a more comprehensive comparison of the two data sets, done by
projecting the dielectron yield observed with HADES into the DLS acceptance.
The particular geometry of the two-arm setup of DLS \cite{dls1}, combined with its
pair trigger requiring coincident lepton hits in both arms resulted in an acceptance
for low-mass pairs ($M_{ee}<0.2$ GeV/c$^2$) confined to mostly small transverse momenta
($p_{\perp}<0.2$ GeV/c) and large rapidities ($y>0.6$).  Although the overall coverage
in $p_{\perp}$ and $y$ of HADES is much wider, because of its toroidal magnetic field
configuration, we have collected only small statistics for low-mass pairs with
$p_{\perp}<0.1$ GeV/c and $y > 1.8$.  This affects strongly $\pi^0$ Dalitz pairs, but only
weakly the pair yield at masses $M_{ee}>0.2$ GeV/c$^2$.  A direct comparison of the two
data sets hence needs an extrapolation of the HADES pair yield to that particular part
of phase space. The whole procedure entails in addition the conversion of multiplicities,
measured by HADES, into production cross sections, as given by DLS.

The published pair acceptance filter of DLS \cite{dls1} acts in a three-dimensional
(3d) phase space ($M$ - $p_{\perp}$ - $y$) and therefore the extrapolation of
the HADES data to full solid angle was performed in a 3d representation as well.
For practical reasons, this has been done by (1) projecting out 2d slices
($p_{\perp}$ vs. $y$) of the efficiency- {\em and} acceptance-corrected pair yield for
different mass bins, (2) fitting a reasonable 2d function to those projections,
and (3) using the resulting fits to extrapolate in 3d phase space, mass slice by mass
slice, to small $p_{\perp}$ and large $y$.  The sparse statistics resulting from
spreading the data counts in three dimensions forced us to use two mass slices only,
emphasizing the $\pi^0$ and the $\eta$ Dalitz mass regions, respectively.

The 2d function employed to fit the pair mass slices has been inspired by the following
constraints only:

\begin{itemize}
 \item $dN/dy$ is gaussian and symmetric around mid-rapidity ($y_{1/2}=0.68$ at 1 AGeV),
 \item $dN/dp_{\perp}$ has a quasi-thermal behavior,
 \item the limited statistics imposes a small number of fit parameters.
\end{itemize}

\noindent
Consequently the function chosen to fit the acceptance-corrected data matrix was:

\[ 1/p_{\perp} d^2N/dp_{\perp}dy = \exp{[-c_0 - c_1 p_{\perp} - c_2 (y-y_{1/2})^2]} \]

\noindent
An advantage of this approach over that based on comparing acceptance-filtered
dilepton cocktails with the respective data sets is that it makes, apart from the above,
no assumptions about the dilepton sources involved.

\begin{figure}[ht]

\mbox{\epsfig{figure={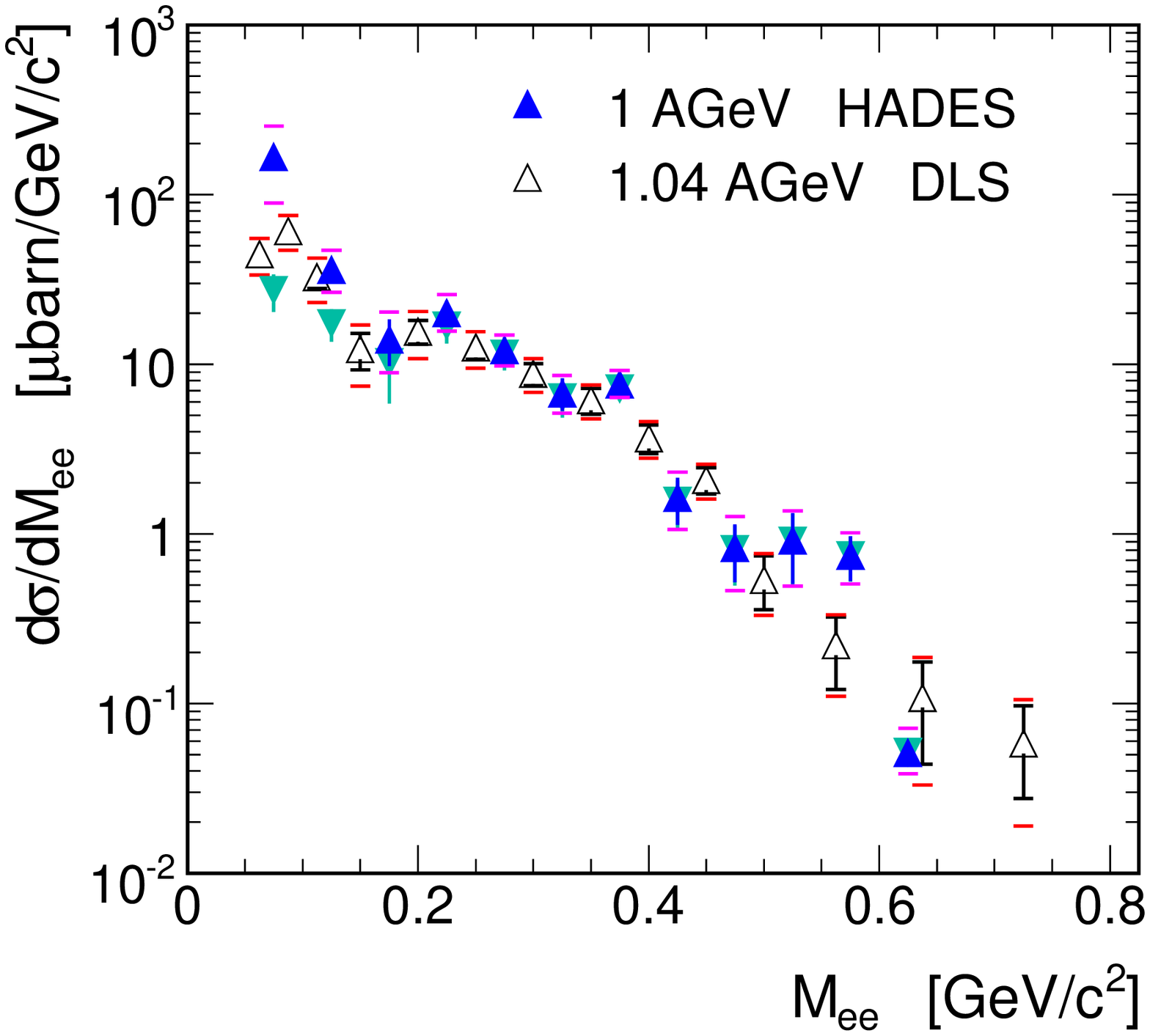}, width=1.\linewidth,height=1.0\linewidth,clip=true}}
\vspace*{-0.0cm}
\mbox{\epsfig{figure={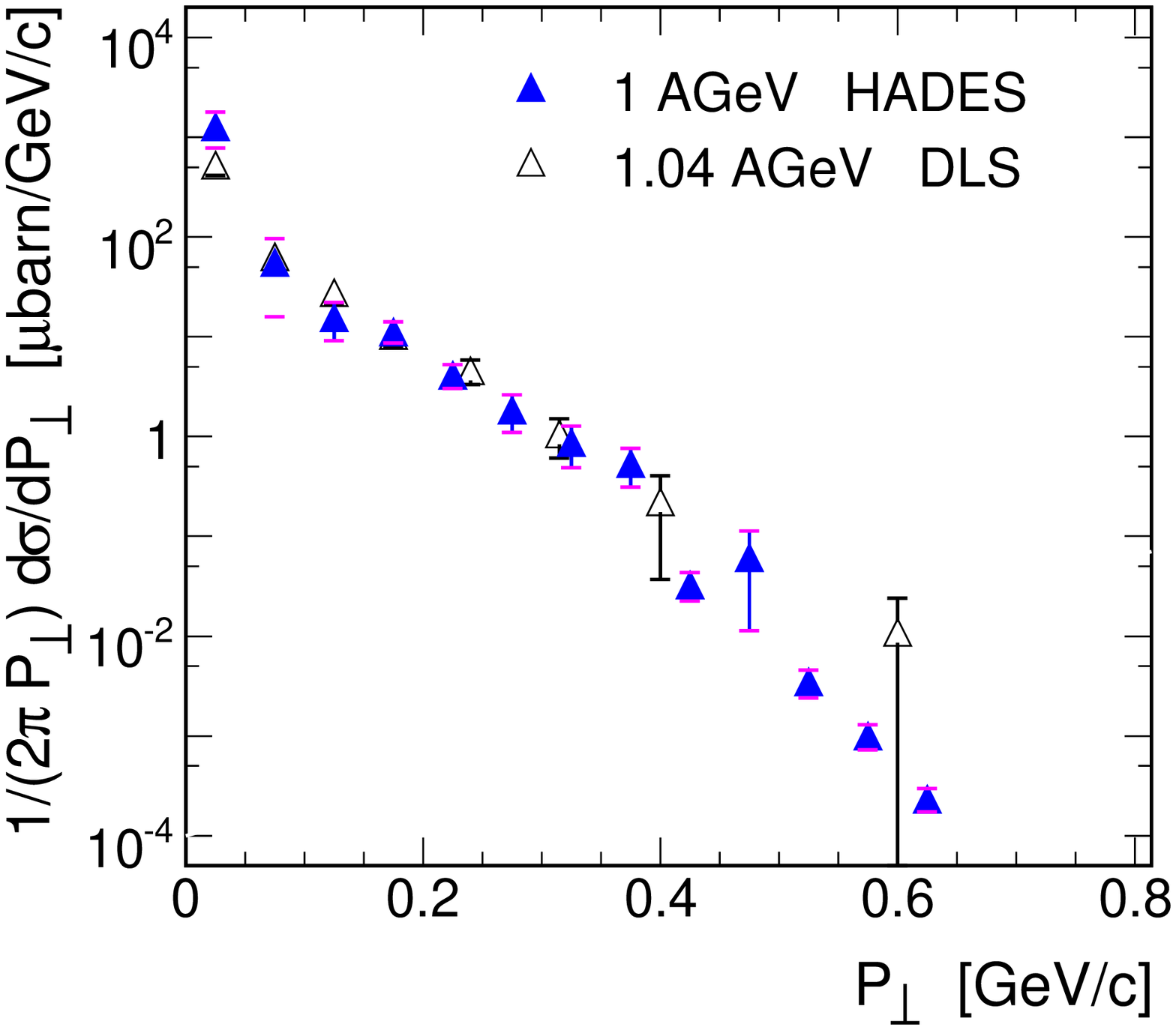}, width=1.\linewidth,height=1.0\linewidth,clip=true}}
\vspace*{-0.5cm}

  \caption[]{
  Direct comparison of the dilepton cross sections measured in C+C at 1~AGeV by HADES
  and at 1.04~AGeV by DLS \cite{dls1, dls2}.  Pair mass distributions (upper frame)
  and pair transverse-momentum distributions (lower frame) are compared within the DLS
  acceptance.  For $d\sigma/dM_{ee}$, both, statistical and systematic errors are shown,
  for $1/(2\pi \, p_{\perp}) \; d\sigma/dp_{\perp}$, only statistical;  for the latter
  data, systematic errors are expected to be large below 0.2 GeV/c \cite{Matis}.
  Overall normalization errors (not shown) are 20\% for the HADES and 30\% for the DLS
  data points.  In the upper frame, the HADES data corresponding to the two fit forms
  discussed in the text are shown as upright ($1/p_{\perp}$) and reverted ($1/p_{\perp}^{2}$)
  full triangles, respectively.
  }

  \label{HADESvsDLS}

\end{figure}

Statistical errors of the data were taken into account in the fit, resulting in
corresponding errors on the fit parameters and accordingly on the extrapolated pair
yield.  The effect of using for the fit a $1/p_{\perp}^{2} d^2N/dp_{\perp}dy$ form
instead was investigated as well and has been taken into account in the systematic
error assigned to the procedure.  The fitted functions were next used to amend the
2d data slices for the yield missing in the region of the acceptance mismatch.
In this operation no attempt was made to compensate for the somewhat larger beam
energy of the DLS experiment, expected to lead to about 5\% (25\%) more $\pi^0$ ($\eta$)
production \cite{taps1, taps2}.  We proceeded by filtering the patched 3d HADES pair
matrix through the DLS pair filter.  The extrapolated part of the reconstructed
yield is reasonably small ($\leq25\%$) in the mass region where we discuss the
excess yield.  At low masses, however, the DLS pair acceptance being quite
different from the HADES acceptance, the correction is sizeable ($\approx90\%$),
with accordingly large systematic errors.

In a final step the HADES multiplicities were converted into cross sections by multiplying
with a total C+C reaction cross section of 900 mbarn and by renormalizing our LVL1 pion
multiplicity (0.53) to its minimum bias value (0.36).  The result of the procedure is given
in Fig.~\ref{HADESvsDLS}, together with the published DLS $d\sigma/dM$ \cite{dls1} and
$1/(2 \pi \, p_{\perp}) \; d\sigma/dp_{\perp}$ \cite{dls2} differential cross sections.
Errors, in particular the systematic effect due to a different choice of the fit
function ($1/p_{\perp}$ vs. $1/p_{\perp}^2$ forms) are indicated as well.  From both
parts of the figure it is apparent that, within statistical and systematic
uncertainties, the HADES and the DLS data are in good agreement, in particular
in the mass region of the excess yield, namely for $M_{ee}=0.15-0.50$ GeV/c$^2$.

\section{Conclusions}

In summary, we report on a measurement of inclusive dielectron
production in C+C collisions at 1~AGeV.  At low masses,
i.e.\ $M_{ee}<0.15$ GeV/c$^2$, the pair yield is in agreement with
the known $\pi^0$ production and decay rates.  For masses of
$0.15$ GeV/c$^2$ $<M_{ee}<0.50$ GeV/c$^2$ it exceeds, however, expectations
based on the known production and decay rates of the $\eta$ meson by
a factor of about $7$.  This excess yield is consistent with that measured
by DLS at 1.04 GeV and a comprehensive comparison of differential
cross sections gives overall good agreement between the two experiments.
The excitation function of the pair yield between 1 and 2~AGeV demonstrates
that the excess scales with bombarding energy like pion production, rather
than like the production of the much heavier $\eta$ meson.

Additional sources associated with the radiation from the early collision phase
($\Delta^{0(+)}\to Ne^+e^-$, $\rho \to e^+e^-$, bremsstrahlung) are clearly
needed to account for the excess observed at $M>0.15$ GeV/$c^2$.  In this context,
our recent studies of pp and pd reactions \cite{hades3} will help by adding
information on dilepton production in elementary reactions, a mandatory input to
any transport calculation.  Indeed, transport models, besides offering a realistic
treatment of the collisions dynamics, also handle the propagation of broad
resonances, related off-shell effects and multi-step processes, all known to play
a crucial role at our bombarding energy.  Better insight is therefore expected once
more refined dynamic calculations become available.

\section{Acknowledgments}

We thank H. W. Matis (LBNL) for valuable clarifications on the DLS experiment.
The collaboration gratefully acknowledges the support by BMBF grants
06MT238, 06GI146I, 06F-171, and 06DR135 (Germany), by the DFG cluster
of excellence {\em Origin and Structure of the Universe (www.universe-cluster.de)},
by GSI (TM-FR1, GI/ME3, OF/STROE), by grants GA ASCR IAA1048304, GA CR 202/00/1668
and MSMT LC7050 (Czech Republic), by grant KBN 1P03B 056 29 (Poland), by
INFN (Italy), by CNRS/IN2P3 (France), by grants MCYT FPA2000-2041-C02-02
and XUGA PGID T02PXIC20605PN (Spain), by grant UCY-10.3.11.12 (Cyprus),
by INTAS grant 03-51-3208 and by EU contract RII3-CT-2004-506078.


\end{document}